\documentclass[twocolumn,floats,prb,aps,showpacs]{revtex4}
\usepackage[final]{graphicx}
\usepackage[dvipsnames]{color}
\usepackage{dcolumn}
\usepackage{hhline}
\DeclareGraphicsExtensions{.pdf,.eps}

\begin{document}

\title{ Electron-phonon coupling in the $C_{60}$ fullerene within the many-body $GW$ approach}

\author{Carina Faber$^{1,2}$, Jonathan Laflamme Janssen$^{3}$, Michel C\^{o}t\'{e}$^3$, 
E. Runge$^{2}$, X. Blase$^1$ }

\affiliation{ $^1$Institut N\'{e}el, CNRS and Universit\'{e} Joseph Fourier,
B.P. 166, 38042 Grenoble Cedex 09, France. \\ 
$^2$Institut f\"{u}r Physik, Technische Universit\"{a}t Ilmenau, 98693 Ilmenau, Germany. \\
$^3$ D\'epartement de physique et regroupement qu\'eb\'ecois sur les Mat\'eriaux de Pointe (RQMP),
Universit\'e de Montr\'{e}al, C.P. 6128, Montr\'{e}al, Qu\'{e}bec, Canada H3C 317. }

\date{\today}

\begin{abstract}
We study the electron-phonon coupling in the $C_{60}$ fullerene within the first-principles $GW$ 
approach, focusing on the lowest unoccupied $t_{1u}$ three-fold electronic state which is relevant 
for the superconducting transition in electron doped fullerides. It is shown that the strength of 
the coupling is significantly enhanced as compared to standard density functional theory calculations 
with (semi)local functionals, with a  48$\%$ increase of the electron-phonon potential $V^{ep}$
with respect to the LDA value. The calculated $GW$ value for the contribution from the $H_g$ modes 
of 93 meV comes within 4$\%$ of the most recent experimental values. The present results call for a 
reinvestigation of previous density functional based calculations of electron-phonon coupling in 
covalent systems in general.
\end{abstract}

%%%%  71.15.Qe 	Excited states: methodology        (in general)
%%%%  71.20.Tx 	Fullerenes and related materials; intercalation compounds 
%%%%  74.20.Fg 	BCS theory and its development     (in Superconductivity)
%%%%  74.20.Pq 	Electronic structure calculations  (in Superconductivity)

\pacs{ 71.15.Qe, 71.20.Tx, 74.20.Fg, 74.20.Pq }

\maketitle

%%%%%%%%%

\section{Introduction}
Electron-phonon coupling in molecular systems is at the heart of several important 
physical phenomena, including the mobility of carriers in organic electronic devices, 
\cite{Gosar66,Coropceanu07,Fratini09,Ortmann09}
the dissociation of excitons at the donor/acceptor interface in organic photovoltaic 
cells, \cite{Tamura08} or the superconducting transition in molecular solids, from the 
most famous fulleride case \cite{Hebard91,Gunnarsson97,Ganin08} to the recent alkali doped picene. 
\cite{Mitsuhashi10}
Even though the interplay between phonon-mediation and electronic correlations is still 
being discussed to better rationalize the superconducting transition all across the
fullerides family,  \cite{Tosatti} the magnitude of the electron-phonon coupling (EPC)
in $C_{60}$
has been the subject of numerous theoretical and experimental studies since the early 
90s in order to evaluate in particular the effective phonon-mediated attractive potential 
$V^{ep}$ central to the BCS theory.

%%%
Of particular relevance for electron-doped fullerenes, 
the coupling to the lowest unoccupied molecular orbital (LUMO)
was explored extensively on the basis of various theoretical approaches, 
\cite{Gunnarsson97}
from earlier combinations of semi-empirical and density functional theory 
(DFT) calculations \cite{Varma91,Schluter92,Mazin92} 
to fully first-principles DFT studies.
\cite{deCoulomb92,Faulhaber93,Antropov93,Breda98,Manini01,Saito02,Janssen10,Iwahara10}
%% Further, the EPC matrix elements were obtained either
%% by following the evolution of the Kohn-Sham eigenvalues under distortion 
%% along a vibrational mode, \cite{Antropov93,Breda98,Manini01,Saito02,Janssen10} 
%% or extracted from the relaxation energy of the molecule upon charging. 
%% \cite{Varma91,Schluter92,Mazin92,deCoulomb92,Faulhaber93,Saito02} 
%%% Manini 38 meV     
Concerning the contribution of the $H_g$ vibrational modes, values from 38 meV 
to 68 meV were calculated within DFT and (semi)local functionals such as
the local density approximation (LDA), while the $A_g$ modes were found
to provide a much smaller contribution, consistently below about 10 meV.
These calculated energies are significantly lower than the available experimental values
extracted from photoemission (PES) experiments on isolated fullerenes with $V^{ep}$ 
found to extend from 96~meV to 147~meV for the $H_g$ modes contribution,
\cite{Gunnarsson95,Wang05,Hands08} and from 107~meV to 158~meV including
both $A_g$ and  $H_g$ contributions. \cite{Gunnarsson95,Wang05}
%% However, the large spread in the experimental results points out to the 
%% difficulties associated with extracting $V^{ep}$ from the analysis of 
%% PES spectra. \cite{Gunnarsson97}

%%%%%%%%%%%
In recent work, \cite{Saito02,Janssen10,Iwahara10} the  EPC
in $C_{60}$ was revisited using DFT and hybrid functionals.
An important outcome of these studies was a significant increase of $V^{ep}$ with increasing 
percentage of exact exchange within modified B3LYP or PBE functionals. 
This clearly indicates that in such systems, not only the electronic excitation energies, but
also the EPC constants, are very sensitive to the choice of the 
exchange-correlation functional.  Despite the overall better agreement with experiments when 
hybrid functionals are used, it is unclear which amount of exact exchange should 
be used for the fullerenes, or any finite or extended system in general. Further, the evaluation 
of the coupling constant to individual energy levels such as the $t_{1u}$ state in $C_{60}$ relies 
%% in most previous work \cite{Antropov93,Breda98,Manini01,Saito02,Janssen10} 
on the identification of the Kohn-Sham eigenstates and eigenvalues with proper 
electronic quasiparticle states and excitation energies.  From a pragmatic point of view, the 
amount of exact exchange in e.g.  B3LYP 
is adjusted to reproduce ground-state properties of a set of molecular systems, \cite{B3LYP}
but does not guaranty that Kohn-Sham eigenvalues reproduce correctly quasiparticle energies.
For example, the $C_{60}$ Kohn-Sham HOMO-LUMO  
gap is 2.8 eV within the DFT-B3LYP approach. \cite{ShucklaC60,ZhangC60} This is better 
than the 1.6 eV obtained within DFT-PBE, \cite{PBE} but still significantly smaller than the 4.9 eV 
experimental gap in the gas phase. \cite{NIST} 

%% Summary
In the present work, we study the electron-phonon coupling in the $C_{60}$ fullerene
using the first-principles $GW$ approach providing well-defined and accurate quasiparticle
energies within a parameter-free many-body perturbation theory framework. We focus on the 
threefold $t_{1u}$ lowest unoccupied molecular orbital (LUMO) which forms the conducting
states in electron-doped fullerides and thus determines the superconducting properties.
We find that the electron-phonon potential $V^{ep}$ is increased by as much as
48$\%$ as compared to DFT-LDA calculations, bridging the gap with experimental data.
In particular, the contribution from the $H_g$ modes is now found to be within 4$\%$ of 
the two most recent experimental estimates. The present results may invite to reconsider 
previous DFT calculations of the electron-phonon coupling constants involved
e.g. in the study of the superconductivity in molecular or extended systems.

\section{Methodology}

In the $GW$ quasiparticle formalism, \cite{Hedin65,Strinati80,Hybertsen86,Godby88,Onida02} for 
which decades of expertise exist in the case of bulk systems, \cite{Onida02,Aulbur95}  
the exchange-correlation potential is described by a non-local energy-dependent self-energy 
$\Sigma({\bf{r,r'}}|E)$ which can be expressed as follows:

\begin{eqnarray*}
 \Sigma^{GW}({\bf{r,r'}}|E) &=& { i \over 2\pi } \int d\omega \; 
    G({\bf{r,r'}}|E+\omega) { W}({\bf{r,r'}}|\omega) \\
 G({\bf{r,r'}}|\omega) &=& \sum_n \phi_n({\bf{r}}) \phi_n^*({\bf r'}) / (\omega - {\varepsilon}_n \pm i\delta)
   \\
 W({\bf{r,r'}}|\omega) &=& \int dr" \; V^C({\bf{r,r''}}) \epsilon^{-1}({\bf{r'',r'}}|\omega)
\label{sigma}
\end{eqnarray*}

\noindent
where $G$ is the time-ordered Green's function \cite{infinitesimal} and $W$ the dynamically 
screened Coulomb potential built from the bare Coulomb potential $V^C$ and the non-local
inverse dielectric matrix $\epsilon^{-1}$ at finite frequency.
For extended solids, the ``starting" $({\varepsilon}_n,\phi_n)$ eigenstates used to build $G$ and 
the dielectric response are traditionally obtained from a ground-state  DFT calculation with 
(semi)local functionals such as LDA or PBE.

In the case of isolated molecules, the $GW$ approach was recently thoroughly validated on a large set 
of small molecules \cite{Rostgaard10} and larger organic ones such as fullerenes, porphyrins 
\cite{Blase11} or DNA/RNA nucleobases. \cite{Faber11} An excellent agreement with experiment for 
the ionization energies and electronic affinities were obtained through a simple self-consistency 
on the eigenvalues with DFT-LDA eigenstates used as the starting point. \cite{Blase11,Faber11,starting}
In this latter approach, labeled $GW$ in what follows, the HOMO-LUMO gap of gas phase $C_{60}$ was calculated
to be 4.91 eV, \cite{Blase11} in much better agreement with experiment than the DFT-B3LYP Kohn-Sham value.

%%  Technical details
Our calculations are based on a recently developed \cite{Blase11,Blase04} 
gaussian-basis implementation of the $GW$ formalism (the {\sc{Fiesta}} code)
 with explicit treatment of dynamical correlations through contour deformation
techniques.  We start from DFT-LDA eigenstates calculated with the {\sc Siesta} package \cite{Siesta}
and a double-zeta plus polarization (DZP) basis \cite{DZP} for the description of the valence orbitals
combined with standard norm-conserving pseudopotentials. \cite{TM}  As shown below, the resulting 
electron-phonon coupling potentials are very similar to that obtained with all-electron calculations,
\cite{Antropov93,Saito02,Janssen10} at least at the DFT level for which several studies are available.
While $GW$ calculations exploiting DFT eigenstates generated with pseudopotentials represent the most 
common approach \cite{Hybertsen86,Godby88,Onida02}, a specific aspect of the present gaussian-basis 
implementation is that the auxiliary basis described here below has been optimized \cite{Kaczmarski10,Blase11,Faber11} 
to project onto the products of occupied/unoccupied pseudized eigenstates.

The needed two-point operators such as the dynamical and non-local free-electron susceptibility
$\chi_0(\textbf{r},\textbf{r'}|\omega)$, the screened Coulomb potential $W(\textbf{r},\textbf{r'}|\omega)$
and the self-energy operator $\Sigma^{GW}(\textbf{r},\textbf{r'}|\omega)$, are expressed on an auxiliary 
even-temperered gaussian basis  consisting of 4 gaussians per each (\textit{s,p,d})-channel with localization
decay constant (0.2,0.5,1.25,3.2) a.u.  Such a basis was thoroughly tested in the $GW$ study of  
a retinal chromophore, \cite{Kaczmarski10} of fullerenes, porphyrins or phtalocyanines 
\cite{Blase11} and DNA/RNA nucleobases. \cite{Faber11,productbasis}

For numerical accuracy when calculating the correlation contribution to the self-energy,
we first evaluate $\Sigma_c^{GW}(E)$ on a coarse energy grid to get a first estimate of the 
quasiparticle energy, and then recalculate $\Sigma_c^{GW}(E)$ on a fine energy grid around 
this energy to refine our calculated correlation contribution.  We verify as well that 
performing the imaginary-axis integration needed in the contour deformation technique 
(see Ref.~\onlinecite{Blase11}) with 12 gaussian points yields results within 0.1 meV as
compared to a calculation using 20 gaussian points. 

%% \begin{figure}
%% \begin{center}
%% \includegraphics*[width=0.45\textwidth]{fig1}
%% \caption{(Color online) Evolution of the three $t_{1u}$ LUMO eigenstates as a function of the $H_g$(7) phonon 
%% displacement norm ($\alpha$ in \AA) within LDA, $G_0W_0$(LDA) and $GW$. The dots are actual calculations and the
%% straight lines a linear fit of the calculated point. Inset: typical evolution of 
%% $<\phi_n |\Sigma_c^{GW}(E)| \phi_n >$ 
%% as a function of E around the quasiparticle energy (indicated by an arrow). Here $\phi_n$ is the (LUMO+1) 
%% state for $C_{60}$ deformed along the $H_g$(7) mode with amplitude $\alpha=0.25~\AA$. }
%% \label{fig1}
%% \end{center}
%% \end{figure}

%%%
Following the results of Ref.~\onlinecite{Janssen10}, we use the relaxed structure and phonon 
eigenmodes generated within the DFT-B3LYP approach and a 6-311G(d) basis. \cite{modes}  This
approach was shown to yield the best eigenfrequencies as compared to Raman experiments. 
\cite{notegw,gaussian} 
The EPC matrix elements are evaluated using a direct frozen-phonon technique.  
Namely, we deform the molecule along the $\vec{\textsl e}_{\nu}$ vibrational eigenvectors with 
typical amplitudes of 0.05~\AA\  and compute the slope
$\; (\vec{\textsl e}_{\nu} \cdot \vec{\nabla}) \varepsilon_i$
of the variation with respect to the deformation amplitude of the DFT Kohn-Sham eigenvalues,
and further of the $GW$ quasiparticle energies.
We verify that we remain in the linear regime as confirmed by the small value of 
the error on the regression coefficient within the fitting procedure. 
Group theory analysis shows that the ($t_{1u} \otimes t_{1u}$) direct product projects only onto 
the non-degenerate $A_g$ modes and the five-fold  $H_g$ vibrational modes, significantly reducing 
the number of matrix elements to be calculated. It remains that ten modes can contribute 
to the coupling, so that a very large number of $GW$ calculations are needed.  

To conclude this methodology section, we note that in the present approach based
on frozen-phonon techniques where atoms are explicitly displaced, the calculated
electron-phonon coupling potentials may be subject to errors related to the use
of localized basis (``Pulay errors"). This issue was previously explored at the DFT
level by comparing localized-basis and planewave calculations showing small
differences (see Supplementary Materials, Ref.~\onlinecite{Janssen10}). In the present 
case of $GW$ calculations, we verify here below that increasing both the size of the DFT 
and auxiliary basis, and taking more diffuse auxiliary orbitals, hardly changes the
value of the coupling constants, suggesting again small errors related to the use of
finite atom-centered bases.

\section{Results}

%%%
Our results are gathered in Table I where we provide the EPC potential
contribution $V^{ep}_{\nu}$ for each of the ten relevant modes, including their degeneracy, 
namely:

\begin{eqnarray*}
 V^{ep}_{\nu} =  { g_{\nu}  \over M \omega_{\nu}^2 }
\sum_{i,j=1}^{3} { | < \phi_i | (\vec{\textsl e}_{\nu} \cdot \vec{\nabla}) V^{SCF} | \phi_j > |^2
   \over g_{t1u}^2 }
\end{eqnarray*}

\noindent where $(\vec{\textsl e}_{\nu} \cdot \vec{\nabla}) V^{SCF}$ is the normalized 
variation of the self-consistent potential under distortion of the molecule
along the vibrational mode with index ($\nu$), degeneracy $g_{\nu}$ and frequency 
$\omega_{\nu}$.  The $(i,j)$ indices run over the $t_{1u}$ manifold with $g_{t1u}$=3 degeneracy. 
The above formula is the molecular limit \cite{Schluter92,Antropov93} of the central definition
used in \textit{ab initio} studies of phonon-mediated superconductivity in extended solids.

In the present frozen-phonon approach, \cite{Antropov93,Breda98,Manini01,Saito02,Janssen10} 
the explicit deformation of the molecule diagonalizes the eigenstates with respect to the perturbation, 
leaving only the intraband transitions which, thanks to Hellmann-Feynman theorem, can be calculated 
through the variation of the corresponding energy levels, namely:

$$
 \sum_{i,j=1}^{3} |< \phi_i | (\vec{\textsl e}_{\nu} \cdot \vec{\nabla}) V^{SCF} | \phi_j > |^2
= \sum_{i=1}^{3} | (\vec{\textsl e}_{\nu} \cdot \vec{\nabla}) \varepsilon_i |^2
$$

\noindent with derivatives calculated through finite differences. This connects 
EPC matrix elements and the variation of the electronic energy levels
with respect to vibrational displacements. This approach is similar to that of 
Refs.~\onlinecite{Antropov93,Breda98,Manini01,Saito02,Janssen10} 
but we use both the $GW$ quasiparticle energies and the DFT Kohn-Sham eigenvalues, 
allowing direct comparison.
As an internal accuracy test, following early group symmetry analysis, \cite{Schluter92} 
the trace of an $H_g$ perturbation is zero in the ($t_{1u}$) subspace,
namely: $\; \sum_{i=1}^3  (\vec{\textsl e}_{\nu} \cdot \vec{\nabla}) \varepsilon_i = 0$,
a condition which is  well verified within our DFT and $GW$ calculations.

Our LDA data yield a total 73.4 meV coupling, in good agreement with the 75.8 meV 
all-electron PBE value of Ref.~\onlinecite{Janssen10}.  Comparing to other similar 
calculations, namely extracting the coupling coefficient from the evolution of the 
DFT-LDA Kohn-Sham eigenvalues under molecular distortion, our 65 meV value for the 
$H_g$ modes contribution is also very close to the 68 meV value by Antropov and 
coworkers, \cite{Antropov93}  within a full-potential framework, and the 
67 meV obtained with an all-electron gaussian basis. \cite{Saito02} Consistently 
with early Raman analysis, \cite{Gunnarsson97} all studies agree on the predominance of 
the two high energy $H_g(8)$ and $H_g(7)$ tangential modes, but contributions at lower 
energy such as from the $H_g(2)$ and $H_g(3)$ radial modes are found to be important as well.

\begin{table*}
\begin{tabular}{c|c|c|c|c|c|c|c|c|c}
\hline
\hline
Mode  &  \multicolumn{6}{c}{Theory}  &  \multicolumn{3}{c} {Experiments} \\
\hline
             & $\omega$ (cm$^{-1}$)  &  LDA    &  B3LYP & Hybrids      & $G_0W_0$(LDA)  & $GW$  & Iwahara$^a$ & Hands$^b$  & Gunnarsson$^c $    \\
\hline
   $A_g(1)$  &  496  & 0.5     &  1.2     &  1.2 - 1.7    & 1.2    &   1.0  (107$\%$)    &  &  &     \\
   $A_g(2)$  &  1492 & 7.7     &  10.9     & 10.5 - 12.9   & 13.6    &   15.0    (93$\%$)  &  &  &     \\
  $H_g(1)$   &  265  & 5.1     &   5.8    & 5.3 - 6.0     &  4.4   &  6.4     (27$\%$)   &  &  &    \\
  $H_g(2)$   &  435  & 9.9     &   10.8    & 10.8 - 13.8   & 15.3   &  11.2     (14$\%$)   &  &  &    \\
  $H_g(3)$   & 721   & 9.1     &  11.9     & 11.0 - 16.7   & 12.3    &  13.9   (53$\%$)    &  &  &   \\
  $H_g(4)$   & 785   & 4.2     &  5.2     &   4.2 - 5.3   &  4.7   &   5.6     (36$\%$)    &  &  &   \\
  $H_g(5)$   & 1123  & 4.2     &  5.0     & 5.0 - 6.7     &  4.2   &   5.2    (23$\%$)      &  &  &   \\
  $H_g(6)$   & 1265  & 2.1     &  2.1     & 2.0 - 4.2     &  2.3   &  2.3     (9$\%$)     &  &  &   \\
  $H_g(7)$   & 1442  & 16.9    & 23.0  & 23.0 - 27.7   &  20.0   &  27.6    (63$\%$)    &  &  &   \\
  $H_g(8)$   & 1608  & 13.7    & 17.7   & 17.0 - 19.3  &  15.6  &  20.4     (49$\%$)     &  &  &  \\
\hline
 Total $A_g$ & -     & 8.2     & 12.2  & 12.1 - 14.3     &  14.8  &  16.0   (95$\%$)  & 10.6  &  -    &  11.3  \\
 Total $H_g$ & -     &  65.2   & 81.5  & 80.0 - 96.3   &  78.8   &  92.6   (42$\%$)   & 96.2  & 96.5 & 147.0  \\
\hline
   Total     & -    & 73.4     & 93.7 & 93.7 - 110.7   &  93.6   &  108.6 (48$\%$)    & 106.8  & - & 158.3 \\
\hline
\hline
\end{tabular}
\caption{Calculated electron-phonon coupling contributions to V$^{ep} $  for the  $A_g$ and $H_g$
modes (meV) calculated within LDA, B3LYP (Ref.~ \onlinecite{Janssen10}), DFT with various hybrid
functionals at the same 6-311G(d) level
(column Hybrids with data from Refs.~ \onlinecite{Saito02,Janssen10,Iwahara10}), non-self-consistent
$G_0W_0$(LDA) and $GW$. The percentage of increase as compared to LDA  is indicated in parenthesis. The
experimental data are compiled in the three last columns. \\
$^a$ Ref.~\onlinecite{Iwahara10}, Table V.  \\
$^b$ Ref.~\onlinecite{Hands08}   \\
$^c$ Ref.~\onlinecite{Gunnarsson95}  \\
}
\label{table}
\end{table*}

The central result of the present study is the dramatic 48$\%$ increase of the total coupling 
potential within the $GW$ approach as compared to LDA calculations. $V^{ep}$ is indeed found 
to increase from 73.4 meV (LDA) to 108.6 meV ($GW$).  For the $H_g$ modes, the calculated $GW$ 
value of 92.8 meV agrees well with the two most recent 96.2 meV and 96.5 meV independent experimental 
estimates of Ref.~\onlinecite{Iwahara10} (Table V) and Ref.~\onlinecite{Hands08}, respectively.
Further, the total $GW$ coupling of 108.6 meV is in close agreement with the latest 106.7 meV 
experimental value. \cite{Wang05,Iwahara10,convergency} 
The present results clearly question the accuracy of the EPC calculated within DFT and (semi)local 
functionals.  In view of the remarkable agreement with experiment obtained with the present 
parameter-free $GW$ formalism, one can hope that this approach will improve our understanding 
of phonon-mediated processes in general.

\section{Discussion}

We can now comment on the recent studies performed with hybrids functionals.
Since both the experimental and $GW$  total coupling potential fall within the rather large 
$[$93,111$]$ meV energy range obtained by changing the amount of exact exchange from 20$\%$ to 
30$\%$ in the hybrid DFT approaches, \cite{Saito02,Janssen10,Iwahara10} one could certainly 
build a functional yielding excellent agreement with experiment for this specific $C_{60}$ 
system. However, clearly, the amount of needed exact exchange may vary from one system to 
another (see below).  It is further instructing to compare mode by mode the various approaches.  
Considering e.g. the  $A_g(2)$ and  $H_g(8)$ modes showing
large coupling, it appears that the largest amount of exact exchange tested so far (30$\%$) is not 
enough to reach the $GW$ results.  In contrast, the $GW$
values for the $H_g(n=2,3,5,6)$ modes are well within the hybrid functionals range. This 
suggests that even for a given single molecule, it seems difficult to optimize the amount of exact
exchange so as to reproduce the $GW$ results mode by mode. This observation leads to emphasizing 
the importance of the non-local and dynamical correlation part of the $GW$ self-energy.
In our approach where only the energy levels are updated, but not the wavefunctions,
the differences between DFT-LDA and $GW$ results can only stem from the replacement of
the exchange-correlation functional by the $GW$ self-energy.

%% GW and G0W0
An interesting observation is that a non-self-consistent $G_0W_0$ calculation starting from LDA 
eigenstates (see column 6 of Table I) leads to a coupling constant which is still significantly 
larger than the DFT-LDA value, but smaller than the $GW$ one, and very similar to that of 
the hybrid B3LYP functional. As  emphasized in recent work, \cite{Rostgaard10,Blase11,Faber11} 
in the case of molecular systems, the significantly too small starting LDA gap leads to a large 
overscreening in the standard $G_0W_0$(LDA) approach.  In the present C$_{60}$ case, the 
$G_0W_0$(LDA) gap is found to be 4.4 eV, much better than the 1.6 eV LDA value, but still smaller
than the 4.9 eV experimental and $GW$ values. Qualitatively, this overscreening certainly softens
the variations of the ionic and electronic potential seen by the electrons upon lattice distortion.

Recently, \cite{Lazzeri08} the EPC matrix elements in graphene for the 
${\Gamma}$-$E_{2g}$ and $K$-$A_{1}'$ phonon modes were studied within a non-self-consistent
$G_0W_0$(LDA) approach.  As compared to DFT-LDA calculations, the square of the deformation 
potentials, labeled ${\langle}D^2_{\Gamma}{\rangle}$ and ${\langle}D^2_K{\rangle}$, 
were shown to increase by 41$\%$ and 114$\%$ respectively. \cite{dopedgraphene}
This is consistent with our own results, suggesting that  EPC
matrix elements are significantly affected by the $GW$ correction in both finite and extended systems.
A further important outcome of this study was that, in graphene, the DFT-B3LYP approach yields significantly
too large coupling constants as compared to experiment, in contrast with the present case of fullerenes
where the DFT-B3LYP calculations underestimate the coupling.  This certainly points out to the difficulties
in obtaining hybrid functionals which are accurate for both finite and extended systems. 

%%
%% As a concluding paragraph, it is interesting to comment on the elegant approach consisting in extracting
%% the electron-phonon coupling matrix elements from the Jahn-Teller distortion of the $C_{60}$ molecule
%% under charging. \cite{Schluter92,Lannoo91} This method, relying on calculating differences of energy 
%% between the neutral and charged fullerene, may seem more appropriate to DFT ground-state  calculations. 
%% It is to be reminded however that the extraction of the electron-phonon coupling to the LUMO level
%% relies on a model, namely the assumption that the relaxation energy along a given vibrational mode consists 
%% in the elastic energy contribution, as given by the phonon-frequency in the harmonic regime, plus an electronic 
%% contribution described by the evolution of the LUMO level energy. This ``independant-electron" picture is  

\section{Conclusion}

In conclusion, we have studied using a first-principles $GW$ approach the electron-phonon coupling
strength in the $C_{60}$ fullerene, focusing on the $t_{1u}$ LUMO state of interest to 
the superconducting transition in the fullerides. It is found that within $GW$, the 
electron-phonon potential $V^{ep}$ increases by 48$\%$ as compared to the value 
calculated within DFT and (semi)local functionals such as LDA or PBE. The calculated    
93 meV $GW$ value for the $H_g$ modes contribution comes within 4$\%$ of the two most 
recent experimental estimates. This demonstrates that the present parameter-free approach 
allows a precise determination of the electron-phonon coupling potential in one of the most 
studied molecular system. Beyond the important case of the fullerenes, the present results call for
a reinvestigation of previous DFT-based calculations of the electron-phonon coupling in organic 
systems, and possibly as well in ``covalent superconducting systems" such as in particular MgB$_2$ or 
doped diamond. Similarly, the important phonon-induced renormalization of the electron and hole
band width in organic semiconductors may deserve further inspection beyond previous DFT calculations.

%%%%
\textbf{Acknowledgements.}
C.F. is indebted to the EU Erasmus program for funding. The authors acknowledge
numerous suggestions from V. Olevano and C. Attaccalite. Calculations have been performed on the 
CIMENT platform (Grenoble) thanks to the Nanostar RTRA project and at IDRIS, Orsay (project 100063).
M.C. and J.L.J. would like to acknowledge the support of NSERC and FQRNT.

%%%%%%%%%% REFS


\begin{thebibliography}{150bliography}
\bibitem{infinitesimal}
The infinitesimal $\delta$ factor formally insures that the poles of G fall in the fourth (second)
quadrant respectively for the unoccupied (occupied) single-particle states.
\bibitem{starting}
The same results are obtained by starting from Hartree-Fock-like energy levels, showing the independence
of the $GW$ quasiparticle energies with respect to the starting point (see Ref.~\onlinecite{Blase11}).
\bibitem{DZP}
The DZP basis was shown in Refs.~\onlinecite{Rostgaard10,Blase11,Faber11} to yield quasiparticle
energies to within 0.1 eV as compared to a much larger TZDP basis.
\bibitem{notegw}
Since the $GW$ formalism is not currently used as a total energy formalism, we adopt the 
usual strategy of performing such calculations on the best available atomic structures.
\end{thebibliography}
\end{document}